\documentclass[12pt]{article}
\usepackage{graphicx}
\newcommand{\beq}{\begin{equation}}
\newcommand{\eeq}{\end{equation}}
\newcommand{\bea}{\begin{eqnarray}}
\newcommand{\eea}{\end{eqnarray}}

\renewcommand{\a}{\alpha}

\begin{document}
\begin{titlepage}
\begin{flushleft}
       \hfill                      {\tt hep-th/0504xxx}\\
       \hfill                       FIT HE - 05-02 \\
       \hfill                       Kagoshima HE-05-1 \\
\end{flushleft}
\vspace*{3mm}
\begin{center}
{\bf\LARGE Meson mass and confinement force \\

 \vspace{.2cm}
driven by dilaton}

\vspace*{5mm} \vspace*{12mm} {\large Iver Brevik \footnote[1]{\tt
iver.h.brevik@ntnu.no}, Kazuo Ghoroku\footnote[2]{\tt
gouroku@dontaku.fit.ac.jp} and Akihiro Nakamura\footnote[3]{\tt
nakamura@sci.kagoshima-u.ac.jp}
}\\
\vspace*{2mm}

\vspace*{2mm}

{\large ${}^{*}$
Department of Energy and Process Engineering, Norwegian University of Science and Technology,
N-7491 Trondheim, Norway}\\
\vspace*{3mm}
{\large ${}^{\dagger}$Fukuoka Institute of Technology, Wajiro,
Higashi-ku}\\
{\large Fukuoka 811-0295, Japan\\}
\vspace*{3mm}
{\large ${}^{\ddagger}$
Department of Physics, Kagoshima University, Korimoto 1-21-35}\\
{\large Kagoshima 890-0065, Japan\\}
\vspace*{3mm}
%\vspace*{4mm}
%{\large ${}^{\ddagger}$Department of Physics, Kyushu University, Hakozaki,
%Higashi-ku}\\
%{\large Fukuoka 812-8581, Japan\\}

\vspace*{10mm}
\end{center}

\begin{abstract}
Meson spectra given as fluctuations of a D7 brane are studied
under the background driven by the dilaton. This leads to a dual
gauge theory with quark confinement due to the gauge condensate.
We find that the effect of the gauge condensate on the meson
spectrum is essential in order to make a realistic hadron spectrum
in the non-supersymmetric case. In the supersymmetric case,
however, only the spectra of the scalars are affected, but they are
changed in an opposite way compared to the non-supersymmetric
case. 

\end{abstract}
\end{titlepage}

\section{Introduction}

In the context of the gauge/gravity correspondence of the
superstring theory \cite{MGW}, an idea to introduce the flavor
quarks has been proposed by Karch and Katz~\cite{KK}. According to
this idea, it becomes possible to study the ``meson'' spectrum by
embedding a D brane in an appropriate 10d background by
identifying mesons as brane fluctuations ~
\cite{KMMW,KMMW2,Bab,ES,SS,NPR}.

In particular, the authors in \cite{KMMW} have given a complete
analysis for the spectrum of the fluctuations on a D7 brane
embedded in AdS$_5\times S^5$. They represent the mass spectra of
mesons formed in a ${\cal N}=2$ SYM theory. In this gauge theory,
however, the potential between quark and anti-quark is Coulomb
like at large distance, so the gauge theory is not in the
confinement phase.
%Meanwhile, at short distance,
%the theory approaches to the conformal theory, and a linear-rising
%potential is observed in this region. %So the mesons in the
%${\cal N}=2$ SYM might be formed by the very short distance force.

\vspace{.3cm} It would be meaningful to extend this analysis to
the case of a background which is dual to a gauge theory with
confinement.  We consider here a background which is deformed by
the dilaton and the axion. In this case, quark confinement is
realized in the sense that a linear-rising potential is found at
large distances. And the QCD string tension in this model is
proportional to the gauge condensate. This point is assured by the
running gauge-coupling constant, which diverges in the infrared
limit. This force should be responsible for forming the meson
states of quark and anti-quark connected by the QCD string. It
would be important to see how this point follows from the meson
spectra.

\vspace{.3cm} Our purpose here is to perform the analysis  from
the gravity side in the background stated above. In a given
background, a D7-brane probe is embedded to add the flavor quark.
Then, by using the embedding configuration of the D7 brane, the
mass spectra of mesons are examined through the fluctuations on
the embedded D7-brane. The analysis is performed first for a
supersymmetric background as given in~\cite{KS2,LT}. In this
background  there is no singularity, and the quark confinement has
been assured for heavy quarks~\cite{KS2,LT} and also for flavor
quarks. Then we extend the same analysis to a non-supersymmetric
case where chiral symmetry is also broken. % for the magnetic condensate case.
%The analysis is also performed for the electric condensate case.

\vspace{.2cm} In section 2, we give the setting of our model and
the embedding of a D7 brane in the supersymmetric background, and
we study the mass spectra of the mesons. In section 3 the analysis
is extended to the non-supersymmetric case.  Summary and discussion are
given in the final section.
%%%%%%%%%%%%%%%%%%%%%%

\section{Background geometry and embedding of a D7 brane}

The D7 brane embedding is briefly reviewed for two types of
backgrounds, supersymmetric and non-symmetric, as given in
\cite{GY}.
%%%%%%%%%%%%%%%% Modified

\subsection{Supersymmetric background}

We consider the $ISO(1,3) \times SO(6)$ symmetric solution given
in ~\cite{KS2,LT} for 10d IIB model. This solution is
supersymmetric and it has no singularity in the bulk, so we can
study the dual gauge theory through the semi-classical approach to
bulk gravity. In the present case, the dual gauge theory for this
background preserves ${\cal N}=2$ supersymmetry. The solution can
be written in the string frame, taking $\alpha'=g_s=1$, as
follows,
 \beq ds^2_{10}=G_{MN}dX^{M}dX^{N}= e^{\Phi/2} \left( {r^2
\over R^2}\eta_{\mu\nu}dx^\mu dx^\nu + \frac{R^2}{r^2} dr^2+R^2
d\Omega_5^2 \right) \ , \label{SUSY-sol} \eeq \beq e^\Phi= \left(
1+\frac{q}{r^4} \right) \ , \quad \chi=-e^{-\Phi}+\chi_0 \ ,
\label{dilaton} \eeq
 where $M,~N=0\sim 9$, $x^{\mu}=X^{\mu}
(\mu,~\nu=0\sim 3)$, $R^4=4 \pi N g_s$ and $q$ is a constant.
$\Phi$ and $\chi$ denote the dilaton and the axion respectively,
and the self-dual five form $F_{\mu_1\cdots \mu_5}$ is given as in
\cite{KS2,LT}. And other field configurations are not used here.
This solution connects two asymptotic geometries, AdS$_5\times
S^5$ and flat space-time, respectively, in IR
($r=0$) limit and UV ($r=\infty$) ~\cite{KS2,LT}.

%%%%%%%%%%%%%%%%%%%%%%%%%%%%%%%%%%

From the holographic context, we give a comment on the form of the
dilaton $e^\Phi$. It represents the running coupling of the dual
gauge theory, and the parameter $q$ is related to the gauge field
condensate via $q/(4\pi N)=\pi^2\langle F_{\mu\nu}F^{\mu\nu}
\rangle$ \cite{LT}. So the parameter $q$ is an important factor
which characterize the vacuum structure of the dual gauge theory,
according to the present model.

%%%%%%%%%%%%%%%%%%%%%%%%%%%%%
\vspace{.3cm} Next, we introduce the flavor quark by embedding a
D7 brane probe which lies in the $\left\{x^{\mu}, X^4\sim
X^7\right\}$ directions. Here we rewrite the 6d geometry as
$\sum_{M=4}^{9}
(dX^M)^2=dr^2+r^2d\Omega_5^2=d\rho^2+\rho^2d\Omega_3^2
+(dX^8)^2+(dX^9)^2$, where $\rho^2=\sum_{M=4}^7(X^M)^2$. Then
$r^2=\rho^2+(X^8)^2+(X^9)^2$. The bosonic part of the brane action
for the D7-probe is
%\beq
%S_{D7} = -\mu_7\int d^8\xi\,
%\sqrt{-\det\left(P[G]_{ab}+2\pi\alpha' F_{ab}\right)}
%+  \frac{(2\pi\alpha')^2}{2} \mu_7\int P[C^{(4)}] \wedge F \wedge F\ ,
%\label{daction}
%\eeq
$$ %\beq
S_{\rm D7}= -\tau_7 \int d^8\xi \left(e^{-\Phi}
    \sqrt{-\det\left({\cal G}_{ab}+2\pi\alpha' F_{ab}\right)}
      +{1\over 8!}\epsilon^{i_1\cdots i_8}A_{i_1\cdots i_8}\right)
$$%% \ ,
\beq
   +\frac{(2\pi\alpha')^2}{2} \tau_7\int P[C^{(4)}] \wedge F \wedge F\ ,
\label{D7-action}
\eeq
where $F_{ab}=\partial_aA_b-\partial_bA_a$.
%${\cal G}=-{\rm det}({\cal G}_{i,j})$, $i,~j=0\sim 7$.
${\cal G}_{ab}= \partial_{\xi^a} X^M\partial_{\xi^b} X^N G_{MN}~(a,~b=0\sim 7)$
and $\tau_7=[(2\pi)^7g_s~\alpha'~^4]^{-1}$ represent the induced metric and
the tension of D7 brane respectively.
And $P[C^{(4)}]$ denotes the pullback of a bulk four form potential,
\beq
C^{(4)} = %e^{\Phi}
\left(\frac{r^4}{R^4} d x^0\wedge d x^1\wedge
d x^2 \wedge d x^3 \right)\ .
\label{c4}
% = \frac{(\rho^2+L^2)^2}{R^4} dx^0\wedge dx^1\wedge dx^2 \wedge dx^3\ .
\eeq
%\vspace{.5cm}
The eight form potential $A_{i_1\cdots i_8}$,
which is the Hodge dual to the axion, couples to the
D7 brane minimally. In terms of the Hodge dual field strength,
$F_{(9)}=dA_{(8)}$ \cite{GGP}, the potential $A_{(8)}$ is obtained as
$A_{i_1\cdots i_8}=-e^{\Phi}\epsilon_{i_1\cdots i_8}$.

\vspace{.5cm}
By taking the canonical gauge, $\xi^a=X^a$, we find an embedding by
solving the equation of motion for the fields $X^8(\xi)$ and $X^9(\xi)$
under the ansatz, $X^9\equiv w(\rho)$ and $X^8=0=F_{ab}$~\cite{GY}.
The stable solution is the supersymmetric one of constant $w$, which
could take an arbitrary value~\cite{GY}. And the induced metric
is written as
\beq
ds^2_8=e^{\Phi/2}g_{ab}dx^adx^b=e^{\Phi/2}
\left(
{r^2 \over R^2}dx^\a dx_\a +
\frac{1}{r^2}R^2 d\rho^2
+\frac{\rho^2}{r^2}R^2 d\Omega_3^2  \right)
\eeq
where $r^2=\rho^2+w^2$ and $g_{ab}$ represents the Einstein frame metric.

Then, the fluctuations of the fields on D7 brane are set as follows,
\beq
 X^9= w+ \phi^9, \quad X^8=\phi^8 \quad{\rm and}~~A_a,
\eeq and the D7 action is expanded in them up to the quadratic
part,
$$%\beq
{\cal L}=-\tau_7~ \left\{e^{\Phi}\sqrt{-\det g_{ab}}\left({1\over 2}g^{ab}
\frac{R^2}{r^2}\sum_{i=8}^9(\partial_a \phi^i \partial_b \phi^i)
+e^{-\Phi}{(2\pi\alpha')^2\over 4} g^{ab}g^{cd}F_{ac}F_{bd} \right)\right.
$$
\beq
\left.-{(2\pi\alpha')^2\over 8}\frac{r^4}{R^4}\epsilon^{\mu\nu\lambda\sigma}
   F_{\mu\nu}F_{\lambda\sigma}+\cdots \right\}.
\label{lag2}
\eeq
%where $F_{ac}=2\pi\alpha'(\partial_aA_b-\partial_bA_a$).

\vspace{.5cm}
%%%%%%%%%%%%%%%%
In the case of $\Phi=0$, the addition of a D7-brane breaks the
supersymmetry to \mbox{${\cal N}=2$}. The light modes coming from
strings connecting the D3-branes and the D7-brane give rise to a
${\cal N}=2$ hypermultiplet in the fundamental representation,
whose field content is two complex scalar fields $\phi^m$ and two
Weyl fermions $\psi_\pm$, of opposite chirality.
%If the D3-branes and the D7-brane overlap then the original
%$SO(6)$ symmetry is broken to
%$SO(4) \times SO(2) \sim SU(2)_R \times SU(2)_L \times U(1)_R$,
%where $SO(4)$ and $SO(2)$ rotate the 4567- and the 89-directions
%respectively. In this case the hypermultiplet is massless
%and the R-symmetry of the theory is $SU(2)_R \times U(1)_R$. The
%fields appear in representations $(j_1,j_2)_s$, where $j_{1,2}$ is the
%spin of $SU(2)_{R,L}$ and $s$ is the $U(1)_R$ charge, in a normalization
%in which the supersymmetry generators transform as $({1\over 2}, 0)_1$.
%With this convention, the two scalars transform in the $({1\over 2},0)_0$
%representation and the two fermions, $\psi_\pm$, are inert under
%\mbox{$SU(2)_R \times SU(2)_L$} and transform with opposite,
%chirality-correlated $U(1)_R$ charges $\mp 1$.
%%%%%%%%%%%%%%%%%

\vspace{.3cm}
Here we consider a non-trivial dilaton, so the supersymmetry is further
broken %and we expect that it is broken
to ${\cal N}=1$ due to the gauge condensate $q$. %or ${\cal N}=0$.
%In our previous analysis, it would be possible to preserve ${\cal N}=1$
%supersymmetry. From this viewpoint, it is interesting to study whether
%the supersymmetry is also preserved in the meson spectra of DBI
%fluctuations.
In the following, we study the meson spectrum to see the effect of
the gauge condensate for the supersymmetric case.

\vspace{.3cm}
\subsection{Scalar field fluctuation}
First of all, we consider the scalars $\phi_i$. They are expressed by the
same linearlized equation of motion,
\beq
 \partial_a\left(e^{\Phi}\sqrt{-\det g_{ab}} g^{ab}
      \frac{R^2}{r^2}\partial_b \delta X \right)=0 \label{scalareq}
\eeq where we abbreviate $\phi_8=\phi_9=\delta X$.
%%%%%%%%%%%%%%%%%%%%%%%%%5
We write $\delta X$ as
\beq
\delta X = \phi(\rho) e^{ik \cdot x} {\cal Y}^{l}(S^3)\ , \label{scalawave}
\eeq
where ${\cal Y}^l(S^3)$ are the scalar spherical harmonics on $S^3$,
which transform in the $(l/2,l/2)$ representation of $SO(4)$
and satisfy
\beq
\nabla^i \nabla_i {\cal Y}^{l} = -l(l+2) {\cal Y}^{l}\ .
\label{defsph}
\eeq
Then equation (\ref{scalareq}) becomes
\beq
\partial_{\rho}^2\phi+\frac{3}{\rho}\partial_{\rho}\phi
+\left(\frac{({mR^2})^2}{(w^2+\rho^2)^2}
-\frac{l(l+2)}{\rho^2}\right)\phi+\partial_{\rho}\Phi\partial_{\rho}\phi=0\ ,
\label{scalar2}
\eeq
where $-k^2=m^2$, which denotes the four dimensional mass.
This equation reduces to the one given in \cite{KMMW} for $\Phi=0$.
The part depending on the dilaton is written as
\beq
  \partial_{\rho}\Phi=-{4q\rho\over (w^2+\rho^2)(q+(w^2+\rho^2)^2)},
\eeq
and it is negligible near $\rho\to 0$ compared to $3/\rho$. So the
infrared behavior of this equation is not changed by this dilaton
term. As for the UV limit, this term is again negligible, and we
obtain the following asymptotic solution,
\beq
  \phi\sim A{\rho^{-l-2}}+B{\rho^{l}} \ .
\eeq
From this, we can see that the conformal dimension of this mode
is $\Delta=3+l$ as expected.

So the effect of the dilaton or of $q=\langle F_{\mu\nu}^2\rangle$ would be seen
in the eigenvalue of the meson mass. For $q=0$, the analytic solution
has been given in \cite{KMMW} and it will be a good approximation
at $\rho\to 0$ and $\rho\to \infty$ for $q>0$ as mentioned above.
Then we can solve the equation numerically by using the analytic
solution for $q=0$ as the boundary conditions. The numerical results
are shown in  Fig.1 for the mass in  the case of $l=0$.

%%%%%%%%%%%%%%% Fig %%%%%%%%%%%%%%%%%%%%% ->
\begin{figure}[htbp]%[H]
\vspace{.3cm}
\begin{center}
  \includegraphics[width=8.5cm, height=6.5cm]{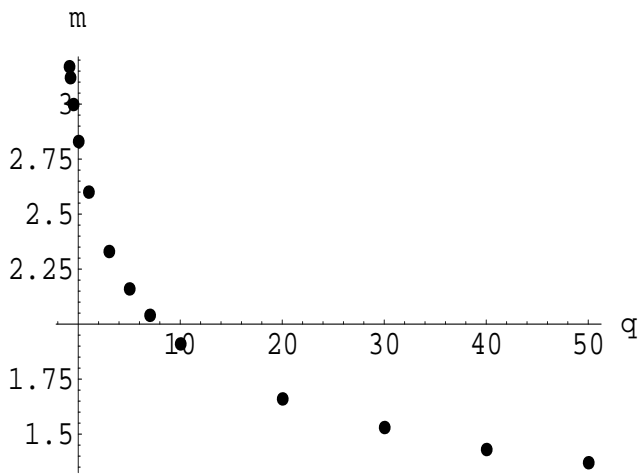}
\caption{Meson mass $m(q)$ vs gauge condensate $q$, where
$m_q=1/(2\pi\alpha')$ and $R=1$.
 \label{susy-mass}}
\end{center}
\end{figure}
%%%%%%%%%%%%%%%%%%%%%%%%%%%%%%%%%%%%%%%%% <-

\vspace{.3cm} The value of the mass decreases with increasing $q$.
This is understandable from the viewpoint of the string theory. In
the present case, the potential between quark and anti-quark is
linear with the distance between quarks at large distances. And
its tension is proportional to $\sqrt{q}$. Then we expect that
$m^2\propto 1/\sqrt{q}$. As we are now considering the small mass
meson states,  this relation would be only approximate. However,
we can see the qualitative feature of this relation from our
results.

\vspace{.3cm}
\subsection{Gauge field fluctuations}

The equations of motion for the gauge fields on the D7-brane,
which follow from the action (\ref{lag2}), are independent of $\Phi$.
They are given as
\beq
\partial_a(\sqrt{-\det g_{cd}} \, F^{ab})-
\frac{4\rho(\rho^2+w^2)}{R^4}\varepsilon^{bjk} \partial_j A_k=0 \,,
\label{geom}
\eeq
where $\varepsilon^{ijk}$ is a tensor density (i.e., it takes values
$\pm 1$). %and we index the coordinates as before (see footnote \ref{note}).
The second term comes from the
Wess-Zumino part of the action, proportional to the pullback
of the RR five-form field strength, and is present only if $b$ is one of
the $S^3$ indices.

Then the gauge field fluctuations are not affected by the dilaton.
This seems to be very strange since the constituent quarks of the
meson state would be influenced by the gauge interaction
characterized in terms of the dilaton. But the above equation
implies that the force responsible to make the meson bound state
is not changed from the one which could not confine quarks. And
the confining force, which is responsible to produce the linear
potential between quark and anti-quark, affects only  the scalar
fluctuations.

\vspace{.5cm} In this sense, there are many kind of bound states
in the present model. The quark and anti-quark in the vector
mesons can be separated at long distances from each other since
they are bounded by the Coulomb-like potential at large distance.
But a quark can not be singled out as a free quark state because
it would have an infinite mass. This can be  understood from the
Wilson loop, which is taken in the $t-r$ plane, as follows. The
relevant part of the metric for this Wilson loop is \beq
 ds^2=e^{\Phi/2}\left(-{r^2\over R^2}dt^2+{R^2\over r^2}dr^2\right).
\eeq
Then the action for the $t-r$ Wilson loop is obtained as
\beq
 S_{t-r}={1\over 2\pi}\int dtdr e^{\Phi/2},
\eeq and the value of this action for a unit time represents the
effective quark mass, $\tilde{m}_q$. It is obtained as
 \beq
 \tilde{m}_q={1\over 2\pi}\int dr e^{\Phi/2}
    ={1\over 2\pi}\int_0^{r_{max}} dr~\sqrt{1+{q\over r^4}} .
\eeq
This implies $\tilde{m}_q=\infty$ for any finite $m_q=r_{max}/2\pi$, which
corresponds to the current quark mass. Then it would be impossible to observe
a single quark. In other words, the quark is confined.

%\section{Fermion mass and supersymmetric spectra}

\vspace{1cm}
%%%%%%%%%%%%%%%%%%%%%%%%%%%%%%%%%%%%%%%%%%%%%%%%%%%%%%%%%%%%%%%%%%%%%%%

\vspace{.3cm}
%%%%%%%%%%%%%%%%%%%%%%%%%

\section{Non-supersymmetric background and meson mass}

Here we consider a non-supersymmetric solution \cite{KS-G,NO,GTU}
which is given without
changing the five form field and eliminating the axion, $\chi=0$, as,
\beq
ds^2_{10}= e^{\Phi/2}
\left(
{r^2 \over R^2}A^2(r)\eta_{\mu\nu}dx^\mu dx^\nu +
\frac{R^2}{r^2} dr^2+R^2 d\Omega_5^2 \right) \ ,
   %\quad A(r)=\left(({r\over r_0})^4-({r_0\over r})^4 \right)^{1/4}
\label{non-SUSY-sol}
\eeq
\beq
 A(r)=\left(1-({r_0\over r})^8 \right)^{1/4}, \quad
 e^{\Phi}=
 \left({(r/r_0)^4+1\over (r/r_0)^4-1} \right)^{\sqrt{3/2}} \ .
   %,  \quad \chi=0 \ ,
\label{dilaton-2}
\eeq
For $\chi=0$, two dilaton solutions $\Phi$ are possible and they
correspond to the magnetic and electric gauge field condensate respectively.
Actually, $e^{\Phi}$ are expanded as
\beq
   e^{\Phi} \sim 1+{q_{\rm{NS}} \over r^4} \; ,  \quad
   q_{\rm{NS}}=\sqrt{6} r_0^4 \; .  \label{condensate}
\eeq
As shown in the previous case, in the context of AdS/CFT, we can
interpret the parameter $q_{\rm NS}$ as
the gauge-field condensate  $\langle F_{\mu\nu}F^{\mu\nu} \rangle
=\langle H_i~^2-E_i~^2 \rangle$.

\vspace{.5cm} These configuration have a singularity at the
horizon $r=r_0$, and the present semi-classical analysis can not
be applied near this point. So we avoid this point in the
following.

\vspace{.5cm}
The D7 brane is embedded as above, and the following brane action is
obtained,
$$ S_{\rm D7-NS}= -\tau_7~\int d^8\xi
{\cal L}_{\rm NS}~~~~~~~~~~~~~~~~~~~~~~~~~~~~~~$$
\beq
=-\tau_7~\int d^8\xi  \sqrt{\epsilon_3}\rho^3 e^{\Phi}
%\left({(r/r_0)^4+1\over (r/r_0)^4-1} \right)^{\sqrt{3/2}}
\left(1-({r_0\over r})^8\right)\sqrt{ 1 + (w')^2 }
\ .
\label{non-SUSY-w9}
\eeq
We %can see that this action shows a repulsive force between D7 and D3 branes,
could expect to find SCSB by solving the equation of motion for $w$,
\beq
  w''+(1+w'~^2)\left\{{3\over \rho}w'+2K_{(1)}\left[
w'(\rho+ww')-(1+w'~^2)w\right]\right\}=0 , \label{equation-NS}
\eeq
\beq
   K_{(1)}={1\over e^{\Phi} A^4}\partial_{r^2}(e^{\Phi} A^4) .
\eeq
%\beq
% \partial_w {\cal L}_{\rm NS}-\partial_{\rho}({\partial {\cal L}_{\rm NS}
%/ \partial {w'}})=0.
%   \label{equation-NS-2}
%\eeq
For some appropriate solution of this equation, we examine the meson mass
by changing the gauge field condensate $q_{\rm NS}$ to see the effects of
the gauge condensate on the meson mass as given in the previous section for
the supersymmetric case.

\subsection{Scalar meson}

With

$$X^8=\phi^8,~X^9=w(\rho)+\phi^9$$
the field equations for $\phi^8$ and $\phi^9$ are
\beq
 \partial_{\rho}^2\phi^8+
    {1\over L_0}\partial_{\rho}(L_0)\partial_{\rho}\phi^8
+(1+w'~^2)
\left[({R\over r})^4{m^2\over A^2}-{l(l+2)\over \rho^2}-2K_{(1)}
\right]\phi^8=0   \label{phi8}
\eeq
\beq
 L_0=\rho^3e^{\Phi} A^4 {1\over\sqrt{1+w'~^2}}, 
  %\quad    K_{(1)}={1\over e^{\Phi} A^4}\partial_{r^2}(e^{\Phi} A^4)
\eeq
and
$$%\beq
 \partial_{\rho}^2\phi^9+
    {1\over L_1}\partial_{\rho}(L_1)\partial_{\rho}\phi^9
+(1+w'~^2)\left[({R\over r})^4{m^2\over A^2}-{l(l+2)\over \rho^2}
     -2(1+w'~^2)(K_{(1)}+2w^2K_{(2)})
\right]\phi^9
$$%\eeq
\beq
 =-2{1\over L_1}\partial_{\rho}(L_1 w~w'K_{(1)})\phi^9
\eeq
\beq
 L_1={L_0\over {1+w'~^2}}, \quad
   K_{(2)}= {1\over e^{\Phi} A^4}\partial_{r^2}^2(e^{\Phi} A^4).
\eeq
In deriving these, we used
\beq
 r^2=\rho^2+(\phi^8)^2+(\phi^9)^2+w^2+2w \phi^9
\eeq
but in the above field equations it is understood as
$r^2=\rho^2+w^2$ since we are considering the linearized equations.

\vspace{.3cm} In order to see the Nambu-Goldstone (NG) mode, which
should be appearing due to the chiral symmetry breaking of the
flavor quark, we rewrite the above equation in terms of polar
coordinates in $X^8-X^9$ plane,
\beq
 X^8=p\sin(\theta), \quad X^9=p\cos(\theta), \quad p=w(\rho)+\delta p~.
\eeq
Here $\theta$ and $\delta p$ are the fluctuations. Then the above equations
are rewritten in the linearized approximation as
$ %\beq
 X^8=w\theta, \quad X^9=w+\delta p~.
$ %\eeq
Then the fluctuations given above are identified as
\beq
 \phi^8=w\theta, \quad \phi^9=\delta p~.
\eeq
Thus the equation for $\theta$, which should be the NG mode, is obtained
by substituting $\phi^8=w\theta$ into Eq.(\ref{phi8}). Then we get
\beq
\partial_\rho^2\theta+\left(2{w'\over w}+
{1\over{ L_0}}\partial_\rho L_0\right)\partial_\rho\theta
+(1+w'~^2)\left[{R\over{r}}^4{{m^2}\over{A^2}}
-{{l(l+2)}\over{\rho^2}}\right]\theta=0    \label{theta}
\eeq

\vspace{.3cm} In order to see the lowest mode of $\theta$, the
s-wave of this mode is studied. Equation (\ref{theta}) is analyzed
in the region of large $\rho$ by using the asymptotic form of
$w(\rho)=m_q+c/\rho^2$. For large $\rho$, then, Eq.(\ref{theta})
is rewritten as \beq
\partial_\rho^2\theta+\left(-4{c\over m_q\rho^3}+
{3\over\rho}\right)\partial_\rho\theta
+{R^4m^2\over{\rho}^4}\theta=0,    \label{theta2} \eeq First of
all, we consider the case of positive and finite $m_q$, where  we
may find the asymptotic form of the renormalizable solution as
\beq
 \theta={1\over \rho^2}+{a_1\over \rho^4}+\cdots ,
 \quad a_1=-\left({c\over m_q}+{R^4m^2\over 8} \right) \label{theta-sol}
\eeq From the second equation of (\ref{theta-sol}), \beq
 m^2=m_0^2-{8\over R^4}{c\over m_q}, \quad m_0^2=-{8a_1\over R^4}.
\eeq
%\vspace{.3cm}
Since ${c/m_q}>0$, $m_0^2$ should be positive and
$m_0^2>{8\over R^4}{c\over m_q}$ due to the positivity of $m^2$.
We thus could get a normalizable mode of positive $m^2$
at some value of $m_q$, then we could
consider a smaller point of $m_q$ where $m^2\to 0$ since $c$ is
finite for $m_q\to 0$. So we can say that $m^2$ decreases with decreasing
$m_q$, but we can't use the above formula near $m^2\sim 0$ since
the approximate equation (\ref{theta2}) is not correct for small $m_q$.
However we expect that the minimum of $m^2$ would be realized at $m_q=0$
as shown below.

\begin{figure}[htbp]%[H]
\vspace{.3cm}
\begin{center}
  \includegraphics[width=11.5cm]{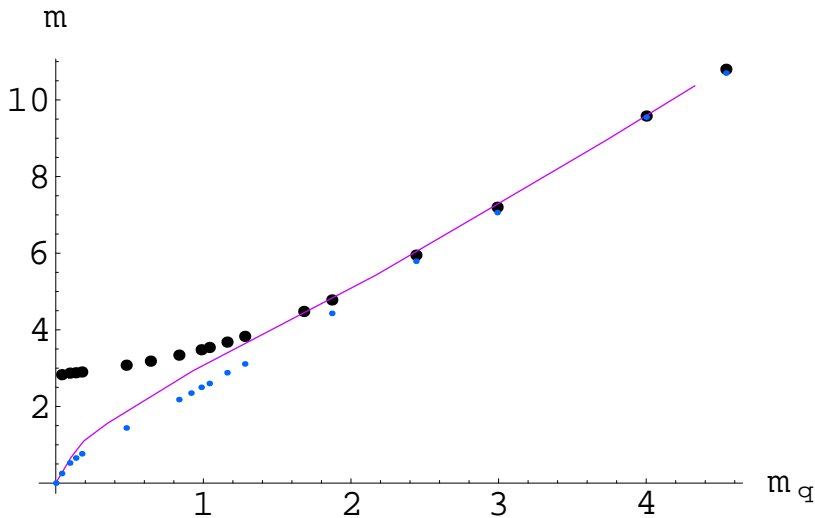}
\caption{Meson mass $m$ for $\phi^8$ (small circle) and $\phi^9$
(large circle) vs $m_q$ for $r_0=1$. The solid curve shows $m$ of
$\phi^8$ for $r_0=2$. Here we take $R=1$.
 \label{non-susy-mass-2}}
\end{center}
\end{figure}

\vspace{.3cm} From Eq.(\ref{theta}), we find a solution of $m=0$
for $l=0$ as $\theta=\theta_0$, which is a constant. Next, we
examine
%Then we consider the case of $m_q=0$ and large $\rho$, and the
%equation is approximated as
%\beq
%\partial_\rho^2\theta-
%{1\over\rho}\partial_\rho\theta
%+{R^4m^2\over{\rho}^4}\theta=0.    \label{theta3}
%\eeq
%For this equation, we obtain the asymptotic normalizable solution as
%\beq
% \theta=1+{b_1\over \rho^2}+\cdots ,
% \quad b_1=-{R^4m^2\over 8} \label{theta-sol-2}
%\eeq
%The second equation of (\ref{theta-sol-2}) is written as,
%\beq
% m^2=m_0^2, \quad m_0^2=-{8b_1\over R^4}.
%\eeq
%In this case, $m^2$ might take some finite value. Then we examine the
%possibility of $m^2=0$ as a minimum value. It is easy to find such a
%solution as $\theta=\theta_0=$constant. Namely, $a_1=\cdots =0$ and
%$\theta_0=1$
%in Eq.(\ref{theta-sol-2}), and this is also an exact solution of
%(\ref{theta}).
%As for
the normalizable condition for this solution, it is given as
\beq
 \int^{\infty} d\rho{1\over \rho}(\phi^8)^2
  \sim \int^{\infty} d\rho{1\over \rho}(m_q+{c\over\rho^2})^2(\theta_0)^2
      < \infty .
\eeq This condition is satisfied only for $m_q=0$. In other words,
there appears a massless mode only for the case of $m_q=0$ and
$c=-\langle\Psi\bar{\Psi}\rangle\neq 0$. In this sense, this zero
mode $\theta$ corresponds to the pion and the lowest massive-mode
of $\phi^9$ would correspond to the sigma meson, which is massive.
We can show these points directly by the numerical estimation
given in  Fig. \ref{non-susy-mass-2}.
%%%%%%%%%%%%%%% Fig %%%%%%%%%%%%%%%%%%%%% ->
%%%%%%%%%%%%%%%%%%%%%%%%%%%%%%%%%%%%%%%%%
In this figure, the lowest mode of $\phi^8$ and $\phi^9$ are
shown. As expected from the above speculation, the mass of
$\phi^8$ approaches  zero with decreasing $m_q$, while the one of
$\phi^9$ arrives at a finite value. On the other hand, both masses
coincide at large $m_q$ and rise linearly. At large $m_q$, $w$ is
almost constant and $w'$ can be neglected, then both equations for
$\phi^8$ and $\phi^9$ approach  the same form when $K_{(2)}$ is
neglected compared to $K_{(1)}$. This point is assured directly by
the numerical estimation. This is the reason why the masses of
$\phi^8$ and $\phi^9$ coincide at large $m_q$.

\vspace{.3cm}
%\subsection{q-dependence}
The massive modes of $\theta$ and $\phi^9$ are obtained by the
numerical analysis. In both cases, we find that the value of the
masses increases with the strength of the gauge condensate $q_{\rm
NS}$ given in (\ref{condensate}). An example for the case of
$\phi^8$ is shown by the solid curve in Fig.\ref{non-susy-mass-2}
for $r_0=2$, which is larger than the case of $r_0=1$. This
behavior is essentially different from the case of the
supersymmetric solution. In the latter case, the mass is
suppressed when the gauge condensate $q$ increases. It would be an
interesting problem to understand this difference from the gauge
theory side, but the problem is left open here.

\section{Summary and discussion}

The background given by (\ref{SUSY-sol}) and (\ref{dilaton}) has
been  extended to the case of of finite temperatures, also with
the  D7 brane embedded in this background \cite{GSUY}. The
fluctuation of the D7 brane in this background would give the
meson spectra at quark-gluon plasma phase. This would lead to
information about the matter properties in the hot universe.

In the context of gauge/gravity correspondence, the
meson spectra were studied by embedding a D7 brane in the background
deformed by the dilaton. We examined both the supersymmetric and
non-supersymmetric background configurations. For the supersymmetric
case, the gauge condensate $q$ affects only the spectrum of the scalar
field and the one of the gauge bosons are independent of $q$. The mass of
the scalar fields decreases with increasing $q$. This implies that the
mass is suppressed when the gauge coupling constant becomes strong.
Then we can suppose that the mechanism to make the bound state in this case
is similar to the case of QED.

On the contrary, in the non-supersymmetric case, the gauge condensate
affects both on the scalar and vector fields. The spectra of the scalars
are examined here and we find that the lowest mass of one of the
scalar fields is zero and other modes are massive when quark mass is zero.
This is understood as the spontaneous chiral symmetry breaking from the
gauge theory side. The masses of the massive modes are increasing with
the gauge condensate $q_{\rm NS}$. Since the string tension is proportional
to $q_{\rm NS}$, the meson, the bound state of quark and anti-quark,
in this case would be connected by the QCD string.

It would be an interesting issue to extend our analysis into the
case of finite temperature. The confinement force driven by the dilaton
is expected to be remaining even at finite temperature \cite{NO,GSUY},
so we will find similar spectra given here for the mesons up to an appropriate
temperature before realizing the complete QGP phase. This will leads
to finding a new phase transition of QCD at high temperature.

In closing, let us make some speculative remarks on the somewhat related topic of Randall-Sundrum brane scenario
\cite{randall99}, assuming that there is only one single brane present.  As is known, all physical processes except
for gravity have to be restricted to lie on the brane. Emission
of gravitons  into the bulk can be imagined to be related to the
interaction between particles on the brane, $\psi
+\bar{\psi}\rightarrow G$. This implies an energy loss equation
for ordinary matter, in the form of a Boltzmann equation, where 
 $C[f]$, the collision term with $f$ the distribution function, can be
written in the form given explicitly in Refs.~\cite{langlois}.

Imagine now that the bulk is not empty but filled with a gas of
gravitons. These gravitons may be absorbed or reflected by the
brane and exert radiation forces on it. Such a process would be
quite analogous to the electromagnetic interaction when a beam of
radiation is incident on a fluid surface, as exemplified in the
classic experiment of Ashkin and Dziedzic \cite{ashkin73}. Actually in
electrodynamics, if one uses  a phase-separated liquid mixture
close to the critical point, whereby the surface tension becomes
very mall, one can manage to create "giant" displacements, of the
order of tenths of micrometers, even with moderate laser powers of
a few hundred of milliwatts \cite{casner01}. The
natural question becomes: Is a brane in a higher-dimensional space
analogous to a surface in electrodynamics, in the sense that it is
 flexible and thus subject to deflections  caused by incident
graviton radiation?

\vspace{.3cm}
\section*{Acknowledgments}

This work has been supported in part by the Grants-in-Aid for
Scientific Research (13135223)
of the Ministry of Education, Science, Sports, and Culture of Japan.

%%%%%%%%%%%%%%%%%%%%%%%
%%%%%%%%%%%%%%%%  References %%%%%%%%%%%%%%%%%%

\newpage
\end{document}